\documentclass[3p]{elsarticle}

\usepackage{amssymb}
\usepackage{amsmath}
\usepackage{lineno}
\usepackage{mathtools}

\begin{document}


\begin{frontmatter}

\title{Oppressed species can form a winning pair in a multi-species ecosystem}

\author[label1]{Attila Szolnoki}
\ead{szolnoki.attila@energia.mta.hu}

\author[label2,label3,label4,label5]{Matja{\v z} Perc}

\address[label1]{Institute of Technical Physics and Materials Science, Centre for Energy Research, P.O. Box 49, H-1525 Budapest, Hungary}
\address[label2]{Faculty of Natural Sciences and Mathematics, University of Maribor, Koro{\v s}ka cesta 160, 2000 Maribor, Slovenia}
\address[label3]{Department of Medical Research, China Medical University Hospital, China Medical University, Taichung 404332, Taiwan}
\address[label4]{Alma Mater Europaea, Slovenska ulica 17, 2000 Maribor, Slovenia}
\address[label5]{Complexity Science Hub Vienna, Josefst{\"a}dterstra{\ss}e 39, 1080 Vienna, Austria}

\begin{abstract}
The self-protection of alliances against external invaders is a key concept behind the maintenance of biodiversity in the face of natural selection. But since these alliances, which can be formed by different numbers of competitors, can also compete against each other, it is important to identify their strengths and weaknesses. Here, we therefore compare the vitalities of two two-species alliances whose members either beat each other mutually via a bidirectional invasion or they exchange their positions during an inner dynamics. The resulting four-species model shows rich behavior in dependence on the model parameter $p$, which characterizes the inner invasions, and $\beta$, which determines the intensity of site exchanges. In the low $p$ and the large $p$ limit, when the inner invasion becomes biased, three-member rock-scissors-paper-type solutions emerge, where one of the members is oppressed by having the smallest average concentration due to heterogeneous inner invasion rates. Interestingly, however, if we allow a more intensive site exchange between the oppressed species, they can morph into a winning pair and dominate the full parameter plane. We show that their victory utilizes the vulnerability of the rival alliance based on cyclic dominance, where a species can easily fixate a limited-size domain.
\end{abstract}


\end{frontmatter}

\section{Introduction}

Life is a permanent fight between species where the more successful one displaces a weaker competitor, according to the Darwinian natural selection principle. But a huge biodiversity of species embrace us, which fact begs an explanation for the stable coexistence of many competitors in an ecosystem. To crack this enigma we should consider that species should not necessarily compete individually, but they may form larger units in which group members mutually help each other directly or indirectly. This possibility establishes the concept of defective alliance which was studied intensively by several excellent works recently \cite{szabo_pre01b,roman_jtb16,nagatani_srep18,szolnoki_epl15,palombi_epjb20,kim_bj_pre05,park_c18c,mao_yj_epl18,wang_m_csf21}. Importantly, the relation of members within the mentioned formation should not necessarily be friendly, but they can be a predator of each other. A simple example could be a trio of $A$, $B$, and $C$ species where every member attacks another group member, but is also attacked by the third party in a similar way as we learned from rock-scissors-paper game \cite{szolnoki_jrsif14,dobramysl_jpa18,han_xz_amc16,tainaka_ei21,szolnoki_epl20}. In other words, $A$ beats $B$, who beats $C$, but the latter also beats $A$. This close loop of dominances may protect group members against the attack of an external, say, $D$ species. If, for example, $A$ is a prey of $D$, but $C$ is a predator of the latter species, then the original $C \to A$ process can block the aggressor $D$, hence the cyclic dominance can maintain its territory.

Evidently, the mentioned loop can be extended to larger groups where the cycle contains four, five, six, or even more species \cite{szabo_pre08,brown_pre19,avelino_pre14,park_c19b,baker_jtb20,vukov_pre13,avelino_csf22,esmaeili_pre18}. But already a pair of species can also form such an alliance. In the latter case, however, we cannot declare a definite rank between the members, but mutual invasion, or their site exchange can help them to block an external invader. Notably, the emergence of alliances can elevate the competition to a higher level because the mentioned groups can fight against not only an individual species but also against another group. This opportunity raises the natural question whether if we can recognize generally valid concepts which primarily determine whether a group of species, a sort of sub-solution, can beat alternative solutions or not. For instance, it was previously demonstrated that the intensity of inner invasion among trio members could be a decisive factor because the group where the invasion rate is higher can overcome a  trio where similar mechanism works, but the invasion rate is modest. Therefore group members are less effective to block an intruder \cite{perc_pre07b}. The conclusion, however, is more subtle because a slower, but homogeneous rates between the partners could be more effective than a cycle where the average of invasions rates is higher, but their values are too diverse \cite{blahota_epl20}. Furthermore, the size of the groups could also be an important factor, because in general alliances formed by less members could be more successful \cite{de-oliveira_csf22}. In other words, trios, but mostly pairs of species are the champions of these contests.

In this work, we compare the effectiveness of differently organized pairs where they use bidirectional mutual invasions or, alternatively, site exchanges as an inner dynamics to protect their coalition from an external threat. To study the possible consequences we introduce a minimal model where four species are present and each pair use one of the mentioned rules, but the pairs attack the alternative pair in a balanced way. Accordingly, we only need two model parameters which characterizes the inner invasions and the intensity of site exchange. Our principal goal is to explore the resulting evolutionary destinations on the complete $p-\beta$ parameter plane.

\section{Model}

In our minimal spatial model four species, marked by $\it 0, \dots, 3$, are distributed on a square lattice. In our interaction graph there is no empty site, but every node is occupied by one of these species. Accordingly, the actual spatial distribution can be determined by a set of site variables, $s_i = \it 0, \dots, 3$ for every site $i$. The species are divided into two pairs which are functioning differently. While species $\it 0$ and $\it 1$ can invade each other mutually with probability $p$ and $1-p$, the remaining two species may exchange their positions with probability $\beta$ if they are neighbors. Furthermore, we assume that every species is simultaneously a predator and a prey of a species from the other pair. In this way we establish a balanced competition between the pairs.
\begin{figure}[h!]
\centering
\includegraphics[width=7cm]{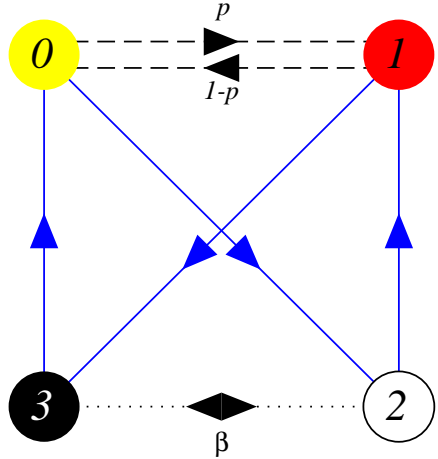}
\caption{Food-web of the four-species model. Arrows with dashed lines stress that invasion can happen between species $\it 0$ and $\it 1$ to both directions with probability $p$ and $1-p$ respectively. Dotted line indicates that neighboring species $\it 2$ and $\it 3$ can exchange their positions with probability $\beta$. Blue arrows with solid lines indicate deterministic invasion between the mentioned pairs in a balanced way. Accordingly, the pair of $\it 0$ and $\it 1$ species may compete with the pair of $\it 2$ and $\it 3$ species and the main question is whether mutual invasion or site exchange is a more effective way to maintain an alliance.}
\label{foodweb}
\end{figure}

The possible interactions are summarized in Fig.~\ref{foodweb}. As the food-web suggests, species $\it 0$ can invade a neighboring species $\it 1$ with probability $p$, while the latter strikes back with probability $1-p$. According to the inner dynamics of the other pair, there is no direct invasion between species $\it 2$ and $\it 3$. If they are neighbors then they exchange their positions with probability $\beta$. The fair competition between the pairs is introduced via deterministic invasion steps, which are denoted by solid blue lines in Fig.~\ref{foodweb}. In particular, when species $\it 0$ meet with species $\it 2$ then the former occupies the position of the latter. On the other hand, a neighboring species $\it 3$ always beats species $\it 0$. Similarly, species $\it 1$ is a predator of species $\it 3$, but is also a prey of species $\it 2$.

To monitor the evolutionary consequence of our model setup we execute Monte Carlo $(MC)$ simulations on the specified $L \times L$ lattice topology, where periodic boundary conditions are applied. Initially each species is distributed randomly hence the starting concentration is $\rho_i = 0.25$ for every $i = \it 0, \dots, 3$ competitors. During an elementary step a nearest-neighbor pair is chosen randomly and if they are occupied by different species then we may execute the invasion or site-exchange according the food-web defined above. To complete a full MC step we need to repeat the elementary step $N = L \times L$ times. As a result, on average every position has a chance to update its status. To avoid the danger of finite-size effects we use large system sizes. In particular, the applied linear size exceeds $L=800$. Notably, when we are close to a discontinuous phase transition point then we may need $L=1600$ linear size to get reliable numerical results. To reach the stationary state we need to apply typically $5\cdot 10^4$ MC steps for relaxation and we average the stationary concentration values over another $10^6$ MC steps to obtain the expected accuracy. For a good statistics every runs are repeated 10 times.

Before presenting our key results, we first discuss the general topology of the food-web summarized in Fig.~\ref{foodweb}. It is easy to notice that for $p>0.5$ values there is a net flow from species $\it 0$ towards species $\it 1$, which establishes a loop among species $\it 0$, $\it 1$, and $\it 3$. In this case there is a symmetry breaking between the members pf the trio, which causes unequal stationary concentrations for the species. Paradoxically it is not species $\it 0$ who suffers from the weak ($p<1$) invasion strength, but species $\it 3$ who is not involved in the modified invasion rate at all. Accordingly, $\rho_3$ is always smaller than $\rho_0 \approx \rho_1$ and this difference becomes significant for intermediate $p$ values. This behavior is frequently called as the ``survival of the weakest'' effect which was first reported by Kei-ichi Tainaka \cite{tainaka_pla95,frean_prsb01,berr_prl09,avelino_pre19b,he_q_pre10,bazeia_csf20,avelino_pre20,avelino_epl21}. It was also shown that this formation of the three species cannot exist if $p$ is close to $1/2$ because the effective inner flow, which serves as a base for the coexistence of the three species, is blocked. Because of the symmetry we can observe a similar viable three-species solution for $p < 1/2$. In this case species $\it 0$, $\it 2$, and $\it 1$ forms a cyclic dominance. But again, due to the unequal invasion rates, species $\it 2$ is oppressed by having a smaller $\rho_2$ value than $\rho_0$ or $\rho_1$ in the stationary state. In our present work both alliances can be present and they may communicate a site-exchange process between the `weaker' species $\it 2$ and $\it 3$.

\section{Results}

\begin{figure}
\centering
\includegraphics[width=15cm]{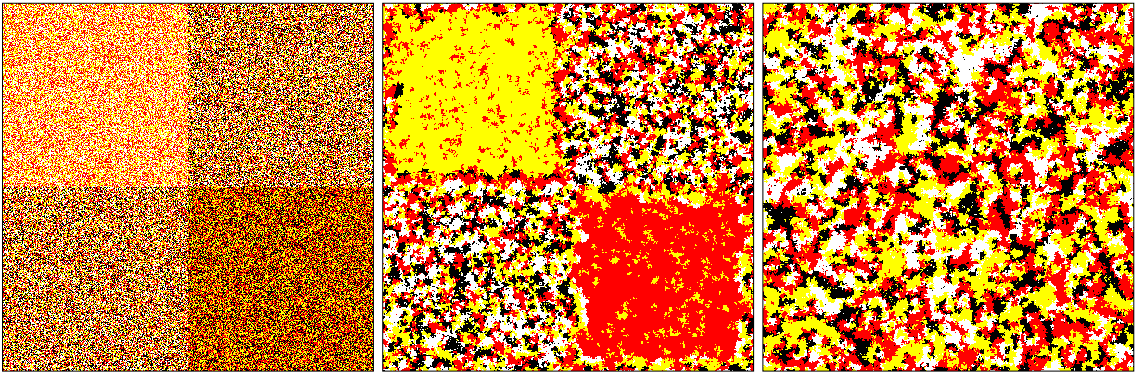}
\caption{Time evolution of spatial distribution of species at $p=1/2$, $\beta=0$. The left panel depicts the prepared initial state where left-up corner contains the mixture of species $\it 0$, $\it 1$, and $\it 2$. The right-down corner is occupied by species $\it 0$, $\it 1$, and $\it 3$. The remaining area is filled by all four species randomly. The color code to mark species is similar to those we used in previous figure. Middle panel shows an early stage of evolution, where the areas belonging to the mentioned three-species solutions start shrinking. In the final state, shown in the last panel, a four-species solution prevails. For a fast simulation we here used $L=400$ and the last two spatial distributions were recorded after 20 and 500 $MC$ steps.}
\label{beta0}
\end{figure}

We first present our results obtained at $\beta=0$ when only invasions are allowed between neighboring species. The parameter region $p \approx 0.5$ is specially interesting, because here neither of the mentioned three-species solutions cannot survive. To illustrate this situation we present a simulation launched from a prepared initial state. This configuration is shown on the left panel of Fig.~\ref{beta0} where we divided the $400 \times 400$ system into four quadrants. In the left-up corner only species $\it 0$, $\it 1$, and $\it 2$ are distributed randomly. Similarly, in the right-down corner species $\it 0$, $\it 1$, and $\it 3$ have a chance to form an alliance. In the remaining space, which contains the right-up and left-down corners, all four species are distributed randomly.

We provide a link to an animation \cite{p05_b0_mix} where the time evolution can be monitored. The key moments, however, are the following. Because of the highly heterogeneous invasion rates within the loops, the portions of `oppressed' species decay very fast and they die out at the early stage of the evolution. In particular, white species $\it 2$ goes extinct in the left-up corner and black species $\it 3$ vanishes in the right-down corner. Theoretically, the remaining two species, yellow species $\it 0$ and red species $\it 1$ are equally strong, but their starting positions are different. Namely, yellow species $\it 0$ occupies the majority of the available area in the first case while red has an initial advantage in the remaining territory of $\it (0, 1, 3)$ loop. Their balanced relation would result in a logarithmically slow coarsening where yellow (red) has a significantly higher chance to prevail \cite{cox_ap86}. But simply there is no opportunity to observe such a destination because the four-species solution invades these frustrated areas faster.

\begin{figure}[h!]
\centering
\includegraphics[width=16cm]{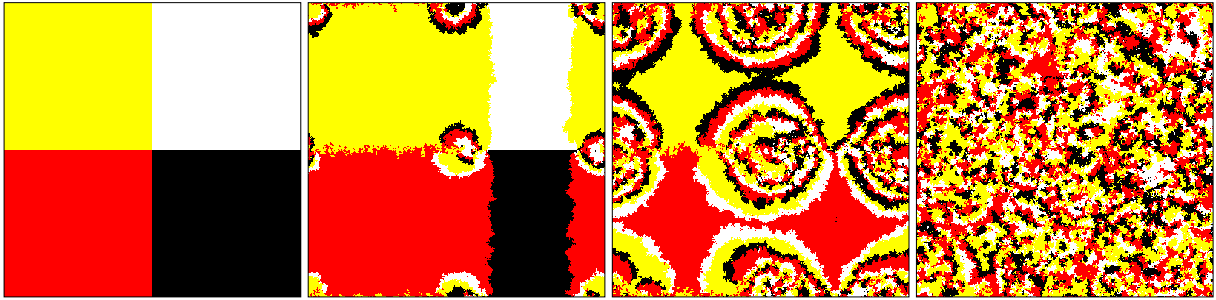}
\caption{Competition of species at $p=1/2$, $\beta=0$. The first panel shows the starting state where each species occupies a quadrant of the available space. When the race is launched, yellow species $\it 0$ starts invading white species $\it 2$. Similarly, red species $\it 1$ attack the territory of black species $\it 3$. This is shown in the second panel, where the interface between the conqueror species becomes noisy due to their balanced bidirectional invasion rates. Notably, $\beta=0$ value prevents mixing the shrinking areas of black and white domains. In parallel, the vortices where all four species meet serve as a birthplace of propagating fronts \cite{szabo_pre99,szabo_pre02,szolnoki_pre10b}. These fronts, shown in the third panel, are the consequence of cyclic dominance among the four species. In the stationary state, shown in the last panel, these fronts are less visible because their collisions break the clear interfaces. The linear system size is $L=400$, where the spatial distributions were recorded after 40, 90 and 500 $MC$ steps.}
\label{waves}
\end{figure}

To help our reader to get further insight about the emerging solution we attached another animation\cite{p05_b0_pure} obtained at the same $p=0.5$, $\beta=0$ parameter values but by using an alternative initial state. In the latter case, as it is illustrated in the first panel of Fig.~\ref{waves}, species are separated first and occupy the four quadrants independently. When we start the simulation the invasion fronts start propagating according to the applied food-web shown in Fig.~\ref{foodweb}. More precisely, yellow species $\it 0$ starts to invade the territory of white species $\it 2$. Similarly, red area of species $\it 1$ grows at the expense of black field of species $\it 3$. The front between yellow and red practically does not move because their relation is balanced. This is strictly true between black and white domains because $\beta = 0$ value prevents any mixing between species $\it 2$ and $\it 3$. The most instructive phenomenon can be observed at vertices where all four species meet. Similarly to the traditional rock-scissors-paper game \cite{szabo_pre99,szabo_pre02} such a vortex is the birthplace of propagating fronts which can invade homogeneous domains easily. The repeating white, red, black, yellow ribbons shown in the third panel of Fig.~\ref{waves} reveal the $\it 0 \to 2 \to 1 \to 3 \to 0$ cyclic dominance among the four species. Notably, such pattern is always a clear sign of a cyclic dominance in a multi-species or in a multi-strategy system \cite{jiang_ll_lnics09,szolnoki_pre10b}
We stress that this pattern remains intact no matter there is an additional invasion-driven interaction between species $\it 0$ and $\it 1$. Of course, the propagating fronts become less visible when they meet and collide that represents the stationary state shown in the last panels of Fig.~\ref{waves} and Fig.~\ref{beta0}.

\begin{figure}[ht]
\centering
\includegraphics[width=8.5cm]{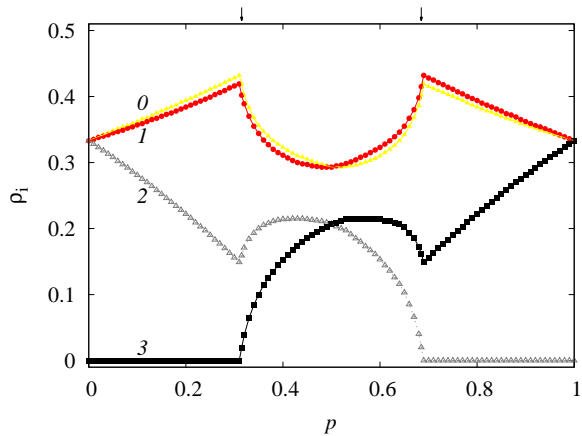}
\caption{The fraction of species as we gradually vary the value of $p$ at $\beta = 0$. If $p$ becomes small or high then the solution formed by four-species is replaced by a three-species loop. The critical points, which sign phase transition and marked by arrows, are at $p=0.31$ and at $p=0.69$. The color is the same we used previously.}
\label{beta_0}
\end{figure}

Summing up our observations, the composition of two infeasible, but overlapping three-member groups can result in a viable four-species group. But what are the limits of the latter solution? To answer this question we explore the full $0 \le p \le 1$ parameter interval and identify the destinations of the evolutionary process. The results are summarized in Fig.~\ref{beta_0}. In this plot we present the fractions of species as we gradually vary the value of $p$ parameter. The results demonstrate clearly that the above mentioned four-species solution is limited to a region around $p=0.5$ point. If $p$ is too low or too high then one of the three-species solutions become dominant. This behavior can be understood in the light of previous observations which revealed when a cyclic dominance could be winning solution. If, for instance, $p$ is large then the invasion from species $\it 0$ toward species $\it 1$ becomes more explicit, which has two consequences. First, the average inner invasion flow among species $\it 0$, $\it 1$, and $\it 3$ is enhanced which always helps a loop to protect against en external intruder more efficiently \cite{perc_pre07b}. From this viewpoint species $\it 2$ can be considered as a lonely external species because there is no circular invasion flow in among species $\it 0$, $\it 1$, and $\it 2$ anymore. Secondly, as $p$ approaches 1, the invasion rates in the $\it 0 \to 1 \to 3 \to 0$ loop become more homogeneous, which makes their alliance more powerful \cite{blahota_epl20}. In the latter case, when a three-member loop is fit enough then it can beat an alliance formed by more partners. This behavior fits again nicely to our expectation about the general effectiveness of defective alliance formed by less members \cite{de-oliveira_csf22}. It is worth noting, however, that the composition of species $\it 0$, $\it 1$, and $\it 2$ is not fair because the latter species is largely `oppressed' and occupy a significantly smaller space compared to the remaining two species. Their balance is restored only in the $p \to 0$ limit.

Evidently, if we use $p<0.5$ values then similar argument can be raised for the $\it 1 \to 0 \to 2$ triplet. Furthermore, there is a general symmetry in our system if $p$ is replaced by $1-p$ then the role of species $\it 0$ and $\it 1$ furthermore the part of species $\it 2$ and $\it 3$ are exchanged. This symmetry can be easily noticed in the functions plotted in Fig.~\ref{beta_0}. For later reference, the `oppressed' member is species $\it 3$ in the last discussed parameter interval.

\begin{figure}
\centering
\includegraphics[width=15cm]{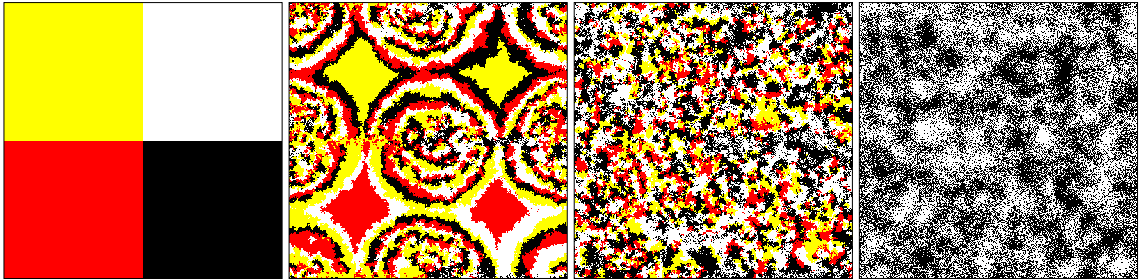}
\caption{Competition of species at $p=1/2$, $\beta=0.06$. When applying the same prepared initial state we used for the mix-free case, the second panel seems to be very similar to the third panel of Fig.~\ref{waves}. But the black and white mixture of species $\it 2$ and $\it 3$ emerges and gradually invades the whole population. The final destination is shown in the last panel. The linear system size is $L=400$, where the spatial distributions were recorded at 100, 220 and 700 $MC$ steps.}
\label{beta_06}
\end{figure}

In the following we allow neighboring species $\it 2$ and $\it 3$ to exchange their positions by using $\beta > 0$ values. To collect a first impression about the consequence of mixing we provide an animation here \cite{p05_b006_pure} and the related representative stages of the evolutionary process are shown in Fig.~\ref{beta_06}. Similarly to the mix-free case, we here also applied a prepared initial state where species are spatially separated first. If we compare the second panel with the third panel of Fig.~\ref{waves}, which were both taken roughly after 100 $MC$ steps, then we can observe very similar patterns. It simply means that the gentle mix, because of the very low $\beta = 0.06$ value, has no detectable consequence in short run. Instead, the outcomes of the more likely invasion steps, which are identical for the compared cases, determine the emerging patterns first. There is, however, a tiny difference, because the black and white mixture of species $\it 2$ and $\it 3$ appears and stands very stably. As the third panel of Fig.~\ref{beta_06} illustrates, the domains of these pair gradually surround the islands of four-species solution. Actually, this step during the evolution already indicates the sad end of the coexistence of four species. As we already stressed, this solution is based on the cyclic dominance of the members. But such solution is always vulnerable in a small domain because the fluctuation of species can easily fix the particular area into a homogeneous state. The latter, on the other hand, could be an easy prey for the tandem of species $\it 2$ and $\it 3$. In this way the latter alliance gradually absorbs disjunct areas previously controlled by the four-species solution.

In other words, the battle between the two-species and four-species solutions does not happen along a propagating frontier where two domains compete openly. Instead, the winning pair emerge randomly in space and their domains slice the original area of four-member alliance into little pieces and killing the rival formation slowly, but surely.

\begin{figure}[h!]
\centering
\includegraphics[width=8.0cm]{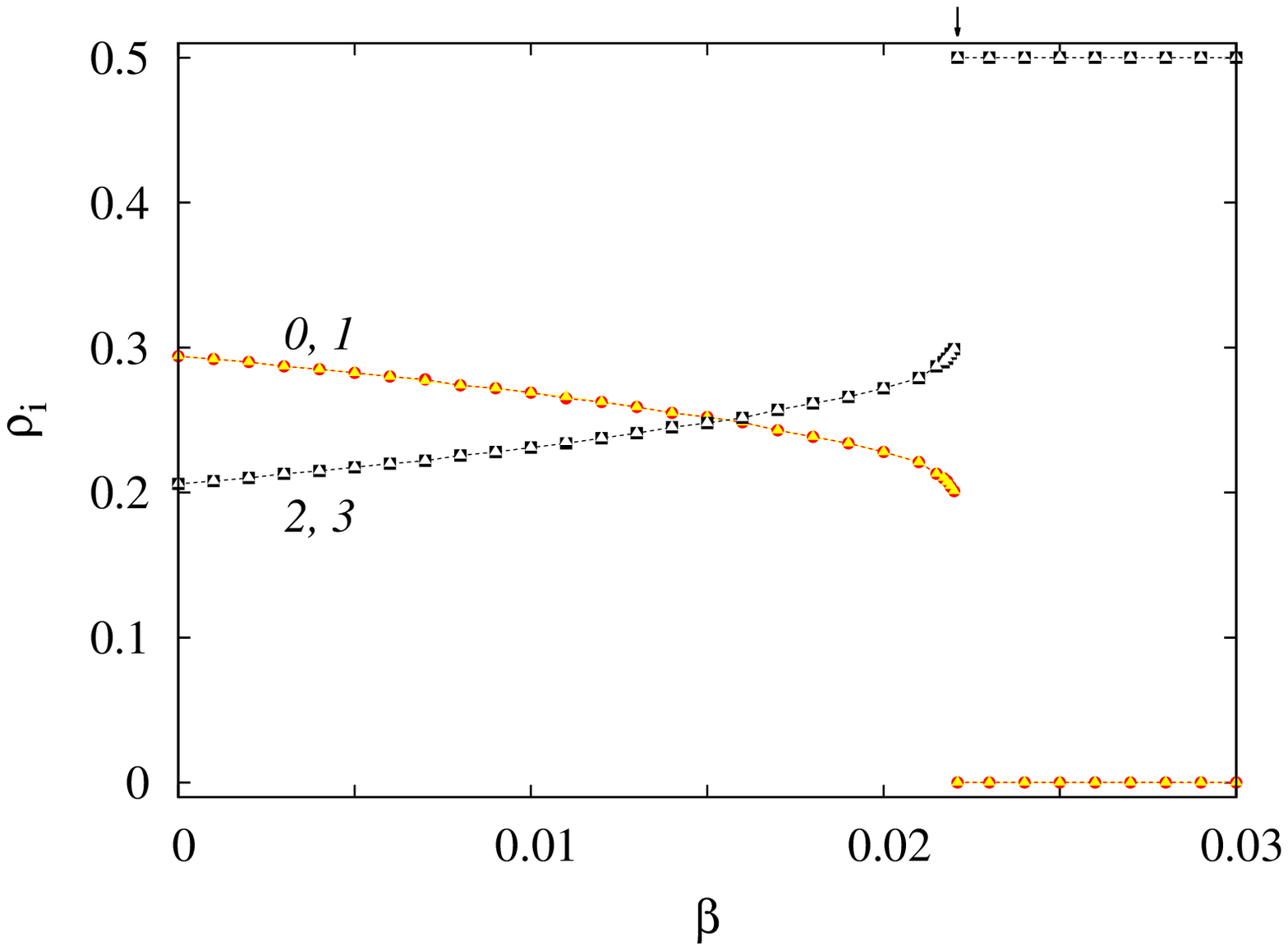}\includegraphics[width=8.0cm]{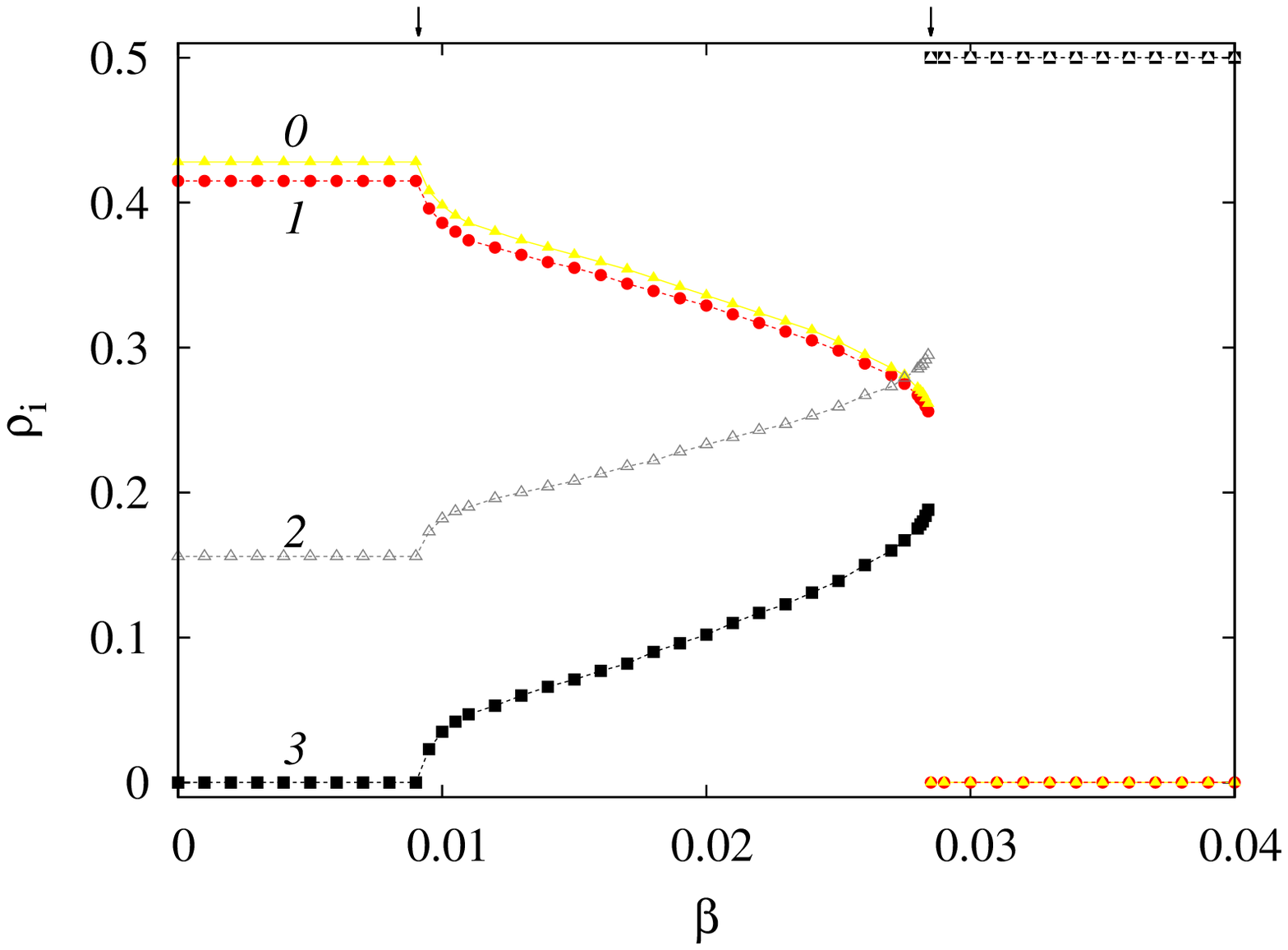}
\caption{Fractions of species as we vary parameter $\beta$ at fixed $p = 0.5$ (left) and at $p=0.3$ (right). In the symmetric $p=0.5$ case the four-species solution is replaced by $\it 2+3$ pair at $\beta = 0.0221$ via a discontinuous phase transition, marked by an arrow. At $p=0.3$ we can observe a continuous transition from a three-species phase to the four-species state at $\beta = 0.0091$ and a discontinuous transition to $\it 2+3$ solution at $\beta = 0.0285$. The applied color code agrees with those used for previous figures.}
\label{p50_30}
\end{figure}

To give a more quantitative account for this transition in Fig.~\ref{p50_30} we plot the stationary concentrations of species as we support the mixing between species $\it 2$ and $\it 3$ by increasing $\beta$. The left panel shows the symmetric case at $p=0.5$. As we can see, the cyclic dominant fours-species solution is replaced by the winning pair of `oppressed' species $\it 2$ and $\it 3$ above a critical mixing rate. Interestingly, already a small value of $\beta$ is capable to change the mentioned species to a winning pair. In other words, it is enough to exchange positions roughly only from 2 of every 100 meets of these species to reach their effective mixing which is capable to beat every other solutions!

Because of the $p=1/2$ value the symmetry of the system is visible in the left panel: the concentration of species $\it 0$ and $\it 1$ always agree, similarly to the equal fractions of species $\it 2$ and $\it 3$. This symmetry is broken for $p = 0.3$, which case is shown on the right panel. Furthermore, in the $\beta \to 0$ limit a three-member cyclic dominant solution emerges, as we already reported in Fig.~\ref{beta_0}. As the cross section shows, the introduction of mixing first destroys the stability of the three-member loop by giving way for the four-member cycle. But increasing the mixing strength further the system terminates into the $\it 2+3$ solution via a discontinuous phase transitions, as we already have seen for the symmetric case.

\begin{figure}[h!]
\centering
\includegraphics[width=8.9cm]{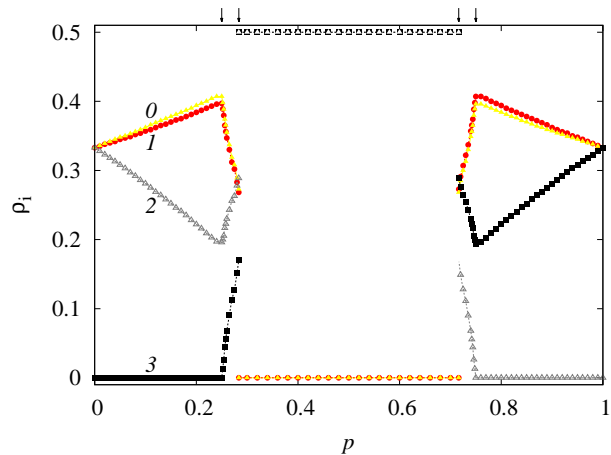}
\caption{The fractions of species as a function of parameter $p$ at fixed $\beta = 0.03$. We can detect four consecutive phase transition at $p$ values $0.25$, $0.2836$, $0.7164$ and $0.75$, marked by arrows. At the extreme (low or high) $p$ values one of the three-member cyclic dominant solutions wins. As we approach the central symmetric case, the system will terminates into a two-member alliance via crossing the four-species phase.}
\label{beta_30}
\end{figure}

If the mixing is strong enough, conceptually similar phase transitions can be observed by only changing the ratio of inner invasion between species $\it 0$ and $\it 1$. A typical behavior can be seen in Fig.~\ref{beta_30}. Similarly to mix-free case, shown in Fig.~\ref{beta_0}, at low $p$ values the cycle of $\it 0+1+2$ species prevails, which is replaced by the four-member loop at $p=0.25$. But this phase is restricted to a narrow range of $p$ because at $p=0.2836$ the pair of $\it 2+3$ becomes the dominant solution via a discontinuous phase transition. Of course, due to the symmetry we already mentioned, we re-enter to the four-species solution at $p=0.7164$, which is replaced by the $\it 0+1+3$ loop in the high $p$ region.

\begin{figure}[h!]
\centering
\includegraphics[width=9.5cm]{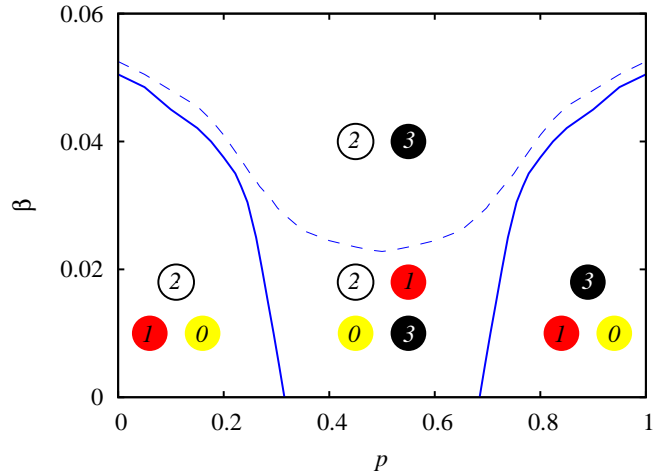}
\caption{Phase diagram on $\beta - p$ parameter plane. In agreement with previous notation, colored bullets mark those species which are present in a specific phase. In the absence or at very low mixing rate between species $\it 2$ and $\it 3$ only three- or four-species alliances prevail which are based on a cyclic dominance. But beyond a critical mixing rate only the two-member solution can survive. Here stable phases are separated by phase transition points. Solid blue lines denote the position of continuous transition. Furthermore, dashed blue line denotes the location of phase transition points which are discontinuous.}
\label{phd}
\end{figure}

The system behavior, more precisely the dominant solutions on the $\beta - p$ parameter plane are presented in Fig.~\ref{phd}, which can be considered as a phase diagram. The message based on this diagram is clear: independently of the inner invasion ratio of species $\it 0$ and $\it 1$, the pair of species $\it 2+3$ can always be successful if their mixing rate is high enough. To reach this evolutionary outcome, the requested $\beta_c$ value is the smallest in the symmetric case around $p = 1/2$, while the most intensive mixing is necessary in the extreme $p \to 0$ and $p \to 1$ limits. In the mentioned limits the inner invasion within the three-species loops become homogeneous, hence the loops are composed by equally strong partners which ensures the maximal fitness to their alliance \cite{blahota_epl20}. The parameter range where four-species solution dominates also becomes narrow in the mentioned limits. This feature can be explained by the fact that the inner invasion flow within the four species turns to be less definite because an alternative invasion flow also emerges. This ambiguity weakens the original concept of defective mechanism which is based on a clear, hence effective, invasion route among group members.

\section{Discussion}

When we are trying to understand biodiversity, our question is often centered on which way is better for a pair of species to protect themselves against external intruders? Either to allow mutual invasions between them, or to allow their mixing in a spatial population? The minimal model where this problem can be studied is a four-species system where pairs using one or the other mentioned strategy to attack the other pair in a balanced way. This system exhibits interesting behavior, where for example three-member and four-member loops can emerge as victors of the evolutionary process at certain values of the control parameters. But the answer to our original question is also revealed clearly. Namely, the spatial mixing of a pair could be a very powerful way to form a defensive alliance, and this effectively so already at a moderate mixing rate. In fact, only a few percentages of their meetings should be a replacement to rule over other solutions on the complete parameter space. Interestingly, the mentioned species are always the `weakest' members of other solutions, because they occupy the smallest area, and hence also reach the smallest concentration levels in comparison with other species. But if we allow them to communicate via mixing, then they become powerful. We also note that bidirectional, or mutual inner invasions could also be vulnerable if they are biased \cite{bazeia_pre19}. In the latter case, one of the partners occupies a larger portion of the available space, and such homogeneity is always a sensitive point of an alliance.

If we monitor the evolution of patterns during the competition between the mixing pair and the alternative cyclic solution, we can observe a specific way that reveals just how the former group prevails. Normally, when two solutions fight for space, their competition is decided at the interfaces that separate the domains of alternative solutions. In the present case, however, the $\it 2+3$ pair wins in a different way. First, their dominance is not visible at the early stage of the evolutionary trajectory because the spatial exchange is less likely in comparison to other invasion steps. Therefore the solution, which is based on the cyclic invasion of group members, can appear faster. On the other hand, the small mixing rate allows a mixed domain to emerge just later in time. But they are robust and thus cannot be swept away by the other solution. Actually, it is best to say that such a pair is not aggressive but simply `patient'. And since the same pair is also member of the rival solution, their tiny domains can emerge anywhere. In this way they can slice the large domains of cyclic solutions into smaller pieces, which then become quarantined and unable to spread. Notably, a small island is always a dangerous time bomb for a solution based on cyclic dominance. That is because the small size is very sensitive to fluctuations of involved species, which are inevitable, and which can thus easily result in a fixation into a homogeneous state. The latter, however, becomes an easy prey for the $\it 2+3$ pair that gradually crowds out the alternative solution.

It is also worth mentioning that a similar type of mixing can emerge rather spontaneously in a mobile system, where the space is not fully occupied, and where species can jump into neighboring empty positions. The chance for species to move freely could also be harmful for a solution which is based on cyclic dominance if the mobility rate is large enough \cite{szabo_pre04,reichenbach_n07,avelino_pre18,mobilia_g16,avelino_csf22,nagatani_jtb19,de-oliveira_pa21}. Lastly, we note that the observed effectiveness of pair mixing could also be valid in broader context of evolutionary game models, because there exist several examples when competing strategies may form a group where their inner relation is conceptually similar to the rock-scissors-paper game \cite{szolnoki_njp14,canova_jsp18,szolnoki_pre17,hauert_s02,szolnoki_csf20b}. To allow neighboring players to exchange their strategies could therefore be a promising research path for finding new, similar types of emerging solutions as reported herein.

\vspace{0.5cm}
This research was supported by the National Research, Development and Innovation Office (NKFIH) under Grant No. K142948 and by the Slovenian Research Agency (Grant Nos. P1-0403 and J1-2457).

\bibliographystyle{elsarticle-num-names}

\end{document}